\documentclass[linenumbers]{aastex631}

\usepackage{multirow}
\usepackage{graphicx}
\usepackage{amssymb, amsmath, amsthm}
\hypersetup{linkcolor=red,citecolor=blue,filecolor=cyan,urlcolor=blue}

\begin{document}
\nolinenumbers

\title{Polarization Calibration of the FAST L-band 19-beam Receiver: I. On-axis Mueller Matrix Parameters}
\correspondingauthor{Tao-Chung Ching}
\email{chingtaochung@gmail.com}

\author[0000-0001-8516-2532]{Tao-Chung Ching}
\altaffiliation{Tao-Chung Ching is a Jansky Fellow of the National Radio Astronomy Observatory.}
\affiliation{National Radio Astronomy Observatory, 1003 Lopezville Road, Socorro, NM 87801, USA}
\affiliation{National Astronomical Observatories, Chinese Academy of Sciences, Beijing 100101, People's Republic of China}

\author{Carl Heiles}
\affiliation{University of California, Berkeley, CA 94720, USA}

\author{Di Li}
\affiliation{National Astronomical Observatories, Chinese Academy of Sciences, Beijing 100101, People's Republic of China}

\author{Timothy Robishaw}
\affiliation{Dominion Radio Astrophysical Observatory, Herzberg Astronomy and Astrophysics Research Centre, National Research Council Canada, PO Box 248, Penticton, BC V2A 6J9, Canada}

\author{Xunzhou Chen}
\affiliation{Research Center for Astronomical Computing, Zhejiang Laboratory, Hangzhou 311100, China}                                                                                                                                                                                                      

\author{Lingqi Meng}
\affiliation{National Astronomical Observatories, Chinese Academy of Sciences, Beijing 100101, People's Republic of China}
\affiliation{School of Astronomy and Space Science, University of Chinese Academy of Sciences, Beijing 100049, People's Republic of China}                                                                                                                                                                                                      

\author{You-Ling Yue}
\affiliation{National Astronomical Observatories, Chinese Academy of Sciences, Beijing 100101, People's Republic of China}
\affiliation{CAS Key Laboratory of FAST, National Astronomical Observatories, Chinese Academy of Sciences, Beijing 100101, China}

\author{Lei Qian}
\affiliation{National Astronomical Observatories, Chinese Academy of Sciences, Beijing 100101, People's Republic of China}
\affiliation{CAS Key Laboratory of FAST, National Astronomical Observatories, Chinese Academy of Sciences, Beijing 100101, China}

\author{Hong-Fei Liu}
\affiliation{National Astronomical Observatories, Chinese Academy of Sciences, Beijing 100101, People's Republic of China}
\affiliation{CAS Key Laboratory of FAST, National Astronomical Observatories, Chinese Academy of Sciences, Beijing 100101, China}

\begin{abstract}
\nolinenumbers

We present the polarization calibration of the 19-beam receiver at 1420 MHz within the full illumination of the Five-hundred-meter Aperture Spherical Telescope from October 2018 to March 2023. We perform spider observations to characterize the on-axis Mueller matrix of the central beam. The calibrated polarization percentage and polarization angle of a source with strong linear polarization emission are about 0.2\% and 0.5$^{\circ}$. 
Several parameters of the central-beam Mueller matrix show time variability from months to years, suggesting relatively frequent polarization calibrations are needed.
We obtain the Mueller matrix parameters of the 18 off-center beams with the combination of on-the-fly observations and spider observations. The polarization calibration provides consistent fractional Stokes parameters of the 19 beams, although the Mueller matrix parameters of the off-center beams are not as accurate as those of the central beam.
The Mueller matrix parameters of the central beam do not show a strong dependence on the {reflector surface}. However, we notice different off-center Mueller matrix parameters between the eastern and western sides of the {reflector surface}. 
We provide average parameters of the 19-beam Mueller matrices which should be applicable to observations {from 2020 to 2022 with several caveats.
After applying the average parameters, on-axis fractional linear polarization measurements $\gtrsim$ 10\% and on-axis fractional circular polarization measurements $\gtrsim$ 1.5\% can be considered high-confidence detections. For sources with weak polarization, timely polarization calibrations using spider observations are required.}
\end{abstract}

\keywords{Astronomical instrumentation (799) --- Polarimetry(1278) --- Single-dish antennas(1460) --- Calibration(2179)}

\section{Introduction} \label{sec:intro}
Polarization is an intrinsic property of electromagnetic waves, and polarization observations of astrophysical sources provide valuable information about the physical conditions and radiative processes of the sources \citep{1996T}. The polarization properties of an astronomical signal are modified by the polarimetric responses of a telescope.
The polarimetric response in the main beam of a radio telescope can be decomposed into on-axis leakage terms and off-axis leakage patterns \citep{2001aHeiles,2001bHeiles}.
For example, the dominant terms of on-axis leakage characterize how incoming on-axis unpolarized signal is converted into a polarized response by the telescope, while the beam squint and squash patterns of off-axis leakage can sample spatial variations of the total-intensity sky distribution within the main beam to induce an instrumentally induced response in the on-axis polarized Stokes parameters.

In radio wavelengths, the polarization of astronomical signal could be about one to two orders of magnitude fainter than total intensity.
Since the leakage of a radio telescope is usually at a similar level, the instrumental polarization would be comparable or even stronger than that of the astronomical signal.
Even for an unpolarized source, the convolution of the angular structure of the source and the squint and squash structures can produce a spurious observed polarization, which is crucial in deriving weak astronomical polarization in the diffuse radio emission from extended objects, e.g., the Zeeman effect in the interstellar medium.  Therefore, polarization calibration needs to be carefully performed.

The on-axis {leakage terms} of a single-dish radio telescope can be described by a Mueller matrix \citep{2002H, 2021RH}.
The Mueller matrix is the transfer function of Stokes parameters between the inputs and outputs of a telescope, accounting for the modifications of polarization by the telescope structure, {reflector surface}, receiver, and electronics system. Since the properties of these elements can change with time, relatively frequent measurements of the Mueller matrix are required. After the determination of the Mueller matrix, the intrinsic Stokes parameters of an astronomical signal can be obtained from the observed Stokes parameters.

The Five-hundred-meter Aperture Spherical Telescope \citep[FAST, ][]{2011Nan} is the world's largest filled-aperture radio telescope, providing sensitive polarization capability.
FAST has a 500-m spherical reflector surface with an active reflector system to form a 300-m parabolic illuminated aperture in real-time moving with the positions of celestial objects \citep{2019Jiang}. 
Because different portions of the reflector surface are illuminated toward different sky positions, the dependence of the Mueller matrix on the zenith angle (ZA) and azimuth angle of FAST needs to be measured.
FAST has a full illumination within ZA = 26.4$^{\circ}$ and is operational up to ZA = 40$^{\circ}$. The 19-beam receiver is placed at the focal point of the reflector surface with incoming radiation only reflecting once before reaching the receiver. This causes the incoming Stokes V signal at the receiver to be reversed from that of the astronomical source, which has been accounted for in this work.
In {tracking} observations at ZAs between 26.4$^{\circ}$ and 40$^{\circ}$, the receiver adopts a backward illumination mode in which the phase center of the receiver is tilted in the direction to the center of the 500-m reflector surface \citep{2013Jin}. This backward illumination mode allows the receiver to see the undeformed surface out of the parabolic aperture instead of seeing the warm ground, eliminating the need of a ground screen. The backward illumination mode thus can severely alter the Mueller matrix of the telescope, and this work focuses on the Mueller matrix at small ZAs.

We have been conducting a long-term study of the polarization performance of FAST since its commissioning phase. 
The preliminary polarization calibration of the central beam had been briefly presented in the commissioning report of the 19-beam receiver \citep{2020Jiang} and the paper reporting the Zeeman effect of the HI narrow absorption line \citep{2022Ching}.
As a series of our study, this work presents the derivation of the on-axis Mueller matrices of the 19-beam receiver within the full illumination of the telescope (ZA $\leq 26.4^{\circ}$) at 1420 MHz for the interests of Galactic HI observations \citep{2018Li}. Based on the on-axis polarization calibration, the study of the beam structures of the full Stokes parameters is presented as the second paper (Chen et al.\ in prep.\ hereafter Paper II).
The frequency dependence of the Mueller matrices over the full bandwidth of the 19-beam receiver and the Mueller matrices at large ZAs will be the third and fourth papers in the future.
The convention used in this work is that the polarization angle, position angle, and azimuth angle are measured from North towards East {following IAU definitions and Stokes $V = {\rm RCP} - {\rm LCP}$ as specified by Commission 40\footnote{DOI: https://doi.org/10.1017/S0251107X00031606} at the 1973 IAU meeting \citep{1974CJ}, where RCP and LCP present the $E$-vector rotating clockwise and counterclockwise as viewed from the transmitter \citep{2018IEEE}. Please see Section 6.2 in  \citet{2021RH} for a more detailed discussion about the definition of Stokes $V$ in radio astronomy.}

\section{FAST L-band 19-beam Receiver} \label{sec:fast}
The 19-beam receiver was developed jointly by Australia's national science agency, the Commonwealth Scientific and Industrial Research Organisation, and the National Astronomical Observatories of the Chinese Academy of Sciences, covering from 1050 to 1450 MHz\footnote{https://www.atnf.csiro.au/technology/receivers/FAST\_Multibeam.html}. The receiver is installed with 19 orthogonal linear polarization feeds followed by a temperature-stabilized noise injection system and low noise amplifiers to produce the $X$ and $Y$ signals of the two polarization paths. The noise injection system consists of a single diode whose signal is split into each beam and polarization. The noise diode employs two selectable power outputs at approximately 1.1 and 12.5 K.

The receiver has been working with backends of multiple commensal observation modes of spectral lines, pulsars, FRB, and SETI.
Currently, only the pulsar backend and spectral-line backend have been commissioned to work.
The spectral-line observation backend can simultaneously record data with a wide bandwidth of 500 MHz and a narrow bandwidth of 31.25 MHz.
The spectral-line wide-band and narrow-band modes were used for all our observations.

\begin{figure}[ht!]
\plotone{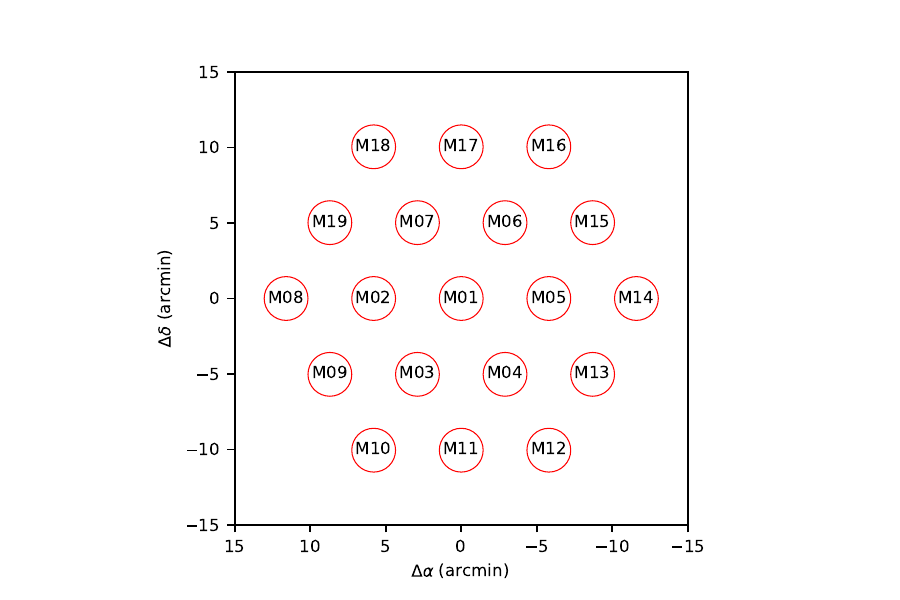}
\caption{The relative positions of the 19 beams to the receiver center at rotation angle $\theta = 0^{\circ}$ in equatorial coordinates with the numbering of the 19 beams from M01 to M19. {Each beam circle has a diameter of the full width at half maximum.}
\label{fig:19beam}}
\end{figure}

For observations at 1400 MHz and ZA $<$ 26.4$^{\circ}$, the system temperature of the 19 beams is between 20 K to 24 K, the aperture efficiency is $\sim 0.63$ for the central beam and $\sim 0.57$ for the off-center beams, and the gain is $\sim 16~{\rm K~Jy^{-1}}$ for the central beam and $\sim 14~\rm{K~Jy^{-1}}$ for the off-center beams \citep{2020Jiang}. The designed angular resolution of the beams at 1420 MHz is 2.9 arcmin with a displacement of 5.8 arcmin between the beams (Figure \ref{fig:19beam}). Please see {Paper II} for the measurements of angular resolutions and structures of the beams. 

\section{Observation and Data Reduction} \label{sec:obs}
Our FAST polarization calibrations of the 19-beam receiver were carried out with spider observations (see \S~\ref{sec:spider}) using the DriftWithAngle mode and on-the-fly (OTF) observations using the OnTheFlyMapping mode.
Table \ref{tab:obs} lists the observations of our polarization calibrations from October 2018 to March 2023, providing {a well-sampled} data set to study the long-term polarization performance of FAST. 
Most of the observations made use of the primary polarization standards 3C48, 3C138, and 3C286 \citep{2013PB} of the Very Large Array (VLA) as our calibrators. The rest of the calibrators were the VLA secondary polarization calibrators\footnote{https://science.nrao.edu/facilities/vla/docs/manuals/obsguide/modes/pol} and the Arecibo polarization calibrators \citep{2022HR} with known strong and stable linearly polarized emission.
The pointing accuracy is typically smaller than 16 arcsec with a root-mean-square pointing error of 7.9 arcsec \citep{2020Jiang}.
The calibration data in 2018 and 2019 were obtained using the polarization commissioning time, and the data from 2020 to 2023 were obtained using regular science projects. 
Except the 3C138 observations on August 19 and 20 in 2019, the ZAs of the observations were smaller than 26.4$^{\circ}$.

The $XX, YY, XY$ and $YX$ correlations from the 19-beam receiver were simultaneously recorded using the narrow-band spectral-line mode of the ROACH2 backend\footnote{
https://casper.berkeley.edu/wiki/ROACH-2\_Revision\_2}, centered at the frequency of the 21-cm HI line. A noise diode signal with a 2 second on/off switching period was injected to calibrate the gain and phase from electronic devices. Before August 2019, we had tried several setups of the noise diode with duty cycles of 10\% or 50\% and sampling rates of 0.1, 0.5, or 1 second. The observations on July 12 and 14 in 2019 employed the low-power noise diode, and the rest of the observations employed the high-power noise diode. After September 2019, the injection of noise diode was optimized for polarization calibration with 10\% duty cycle, 0.1 second sampling rate, and high-power mode. Owing to the relative phase between the two polarization paths, noise diode signal in the cross-products is much stronger than astronomical signal. We thus only used the noise diode off data to generate the $XX, YY, XY$ and $YX$ data of the sources.

Our calibration only made use of the continuum emission of the sources. The radio frequency interference and the HI absorption in the data were removed before the data reduction. The data reduction, including gain and phase calibrations of the two polarization paths, bandpass calibrations, and polarization calibrations was carried out using the IDL RHSTK package written by C.\ Heiles and T.\ Robishaw \citep{2022HR}, which is widely used for the Arecibo and Green Bank Telescope (GBT) data reduction of polarization observations. 
After the calibration, the self- ($XX$ and $YY$) and cross-products ($XY$ and $YX$) were used to obtain the observed Stokes parameters with
\begin{align}
    I_{\rm obs} &= XX + YY \\
    Q_{\rm obs} &= XX - YY \\
    U_{\rm obs} &= 2XY \\
    V_{\rm obs} &= 2YX.
\end{align}

\begin{table}[h!]
\begin{center}
\caption{Observation Parameters}
\label{tab:obs}
\begin{tabular}{ccccccccc}
\hline
\hline
\multirow{2}{*}{Date} & \multirow{2}{*}{Target} & \multicolumn{2}{c}{Pointing Center} &  \multicolumn{2}{c}{Zenith Angle (deg)} & \multicolumn{2}{c}{Azimuth Angle (deg)} & \multirow{2}{*}{Project ID} \\ 
\cline{3-8} & & $\alpha$ (J2000) & $\delta$ (J2000) & start & end & start & end & \\ 
\hline
\multicolumn{9}{c}{Spider Observations} \\
\hline
2018/10/06 & 3C286 & 13:31:08.31 & +30:30:32.3 & \phn4.8 & 14.7 & 185.9& 112.5 & 1005 \\
2018/10/07 & 3C286 & 13:31:08.31 & +30:30:32.3 & \phn4.8  & 14.6 & 185.9& 112.5 & 1005 \\
2019/03/21                     & 3C380                     & 18:29:31.78 & +48:44:46.2 & 23.8 & 25.1 & 192.2& 160.2 & 3053                     \\
2019/03/22                     & 3C48                      & 01:37:41.30 & +33:09:35.1 & \phn9.7  & 13.7 & 216.6& 126.6 & 1005                     \\
2019/07/12                     & 3C48                      & 01:37:41.30 & +33:09:35.1 & \phn9.5  & 14.0 & 215.3& 126.1 & 1005                     \\
2019/07/13                     & 3C48                      & 01:37:41.30 & +33:09:35.1 & \phn9.5  & 14.0 & 215.3& 126.1 & 1005                     \\
2019/07/14                     & 3C48                      & 01:37:41.30 & +33:09:35.1 & \phn9.5  & 14.0 & 215.3& 126.1 & 1005                     \\
2019/08/19                     & 3C138                     & 05:21:09.89  & +16:38:22.1 & 25.1 & 38.9 & \phn74.0& \phn84.9  & 1005                     \\
2019/08/20                     & 3C138                     & 05:21:09.89  & +16:38:22.1 & 25.1 & 38.8 & \phn74.0& \phn84.9  & 1005                     \\
2019/08/21                     & 3C286 & 13:31:08.31 & +30:30:32.3 & 18.0 & 18.0 & 250.2& 109.9 & 1005                     \\
2019/09/05                     & 3C286 & 13:31:08.31 & +30:30:32.3 & 10.0 & 11.5 & 239.2& 117.1 & 1005                     \\
2019/10/23                     & 3C286 & 13:31:08.31 & +30:30:32.3 & 11.5 & 10.0 & 242.9& 120.8 & 1005                     \\
2019/11/23                     & 3C286 & 13:31:08.31 & +30:30:32.3 & 10.7 & 10.7 & 241.2& 118.8 & 1005                     \\
2020/09/12                     & 3C286 & 13:31:08.31 & +30:30:32.3 & \phn9.6  & \phn9.5  & 238.1& 122.1 & PT2020\_0064             \\
2020/11/02                     & J2202+4216                & 22:02:43.29 & +42:16:40.0 & 18.4 & 18.4 & 202.4& 157.6 & PT2020\_0078             \\
2021/02/18                     & 3C273                     & 12:29:06.70 & +02:03:08.6 & 24.5 & 24.5 & 344.8& \phn15.4  & PT2020\_0128             \\
2021/02/22                     & 3C286 & 13:31:08.31 & +30:30:32.3 & \phn7.3  & \phn7.3  & 227.8& 131.9 & DDT2021\_3               \\
2021/09/28                     & 3C48                      & 01:37:41.30 & +33:09:35.1 & 10.6 & 10.6 & 222.2& 137.8 & PT2021\_0130             \\
2021/10/11                     & 3C286 & 13:31:08.31 & +30:30:32.3 & \phn9.3  & \phn9.3  & 237.1& 122.9 & PT2021\_0009             \\
2021/10/18                     & 3C138                     & 05:21:09.89  & +16:38:22.1 & 10.8 & 10.8 & 324.7& \phn35.3 & PT2021\_0074             \\
2021/11/06                     & 3C138                     & 05:21:09.89  & +16:38:22.1 & 11.5 & 11.5 & 319.6& \phn40.3 & PT2021\_0068             \\
2022/09/13                     & 3C48                      & 01:37:41.30 & +33:09:35.1 & \phn9.9  & \phn9.9  & 218.0& 142.0 & PT2022\_0088             \\
2022/09/26                     & 3C286 & 13:31:08.31 & +30:30:32.3 & \phn8.0  & \phn8.0  & 231.9& 129.0 & PT2022\_0001             \\
2023/03/11                     & 3C286 & 13:31:08.31 & +30:30:32.3 & \phn8.0  & \phn8.0  & 231.9& 129.0 & PT2022\_0163             \\ 
\hline
\multicolumn{9}{c}{On-The-Fly Observations} \\
\hline
2020/09/10 & J0854+2006 & 08:54:48.87 & +20:06:30.6 & 13.9 & 12.0 & 291.0&   \phn64.5 & PT2020\_0064 \\ 
2020/09/11 & J0854+2006                & 08:54:48.87 & +20:06:30.6 & 16.3 & 10.6 & 286.8&   \phn59.9  & PT2020\_0064             \\
2020/09/11 & 3C48                      & 01:37:41.30 & +33:09:35.1 & \phn7.9  & 7.7  & 195.9&   172.0 & PT2020\_0064             \\
2020/09/11 & 3C84                      & 03:19:48.16 & +41:30:42.1 & 16.0 & 16.1 & 185.0&   173.3 & PT2020\_0064             \\
2020/09/11 & 3C138                     & 05:21:09.89 & +16:38:22.1 & \phn9.2  & \phn9.1  & 347.6&   \phn\phn8.4  & PT2020\_0064             \\
2021/10/05                     & 3C84                      & 03:19:48.16 & +41:30:42.1 & 15.9 & 16.5 & 181.4&   166.7 & PT2021\_0009             \\
2021/10/06                     & 3C138                     & 05:21:09.89 & +16:38:22.1 & 10.0 & 12.9 & \phn27.2&    \phn47.7  & PT2021\_0009             \\
2021/10/06                     & J0854+2006                & 08:54:48.87 & +20:06:30.6 & 17.3 & 18.7 & 285.3&   \phn76.5 & PT2021\_0009             \\
2021/10/07                     & J0854+2006                & 08:54:48.87 & +20:06:30.6 & 16.9 & 14.7 &  285.9&   \phn70.5 & PT2021\_0009             \\
2021/10/10                     & 3C48                      & 01:37:41.30 & +33:09:35.1 & \phn8.5  & \phn7.6  & 205.4&   175.3 & PT2021\_0009             \\
2022/09/18                     & 3C138                     & 05:21:09.89 & +16:38:22.1 & \phn9.0  & \phn9.6  & \phn\phn0.4&     \phn20.6  & PT2022\_0001             \\
2022/09/19                     & 3C48                      & 01:37:41.30 & +33:09:35.1 & \phn7.6  & \phn8.5  & 176.7&   154.3 & PT2022\_0001             \\
2022/09/20                     & J0854+2006                & 08:54:48.87 & +20:06:30.6 & \phn7.4  & \phn9.9  & \phn42.0&    \phn57.0  & PT2022\_0001  \\          
\hline
\end{tabular}
\end{center}
\end{table}

\section{Polarization Calibration} \label{sec:cal}
Here we provide the details of the polarization calibration including the observational strategies and the pipeline sequence. 
We adopt the authoritative work on the polarization calibration of radio telescopes by \citet{2001aHeiles} as the fundamental reference for our calibration method. Several statements and assertions are made here for a simplified form of Mueller matrix, and the explanations and justifications can be found in \citet{2001aHeiles} and the references within. 

\subsection{Mueller Matrix}
The Mueller matrix represents the conversion of Stokes parameters from input to output by a device. For a radio telescope, the radiation of a source sequentially travels from the telescope dish to the receiver. 
The Mueller matrix hence has two components: the matrix involving the telescope pointing (\textbf{\textit{M}}$_\text{SKY}$) and the matrix involving the receiving system (\textbf{\textit{M}}$_\text{TOT}$). 
The transformation from the intrinsic Stokes parameters of a radio source $(I_{\rm src}, Q_{\rm src}, U_{\rm src}, V_{\rm src})$ to the observed Stokes parameters $(I_{\rm obs}, Q_{\rm obs}, U_{\rm obs}, V_{\rm obs})$ is
\begin{equation}
\begin{bmatrix} 
I_{\rm obs} \\ Q_{\rm obs} \\ U_{\rm obs} \\ V_{\rm obs} 
\end{bmatrix} 
= 
\textbf{\textit{M}}_{\text{TOT}} \cdot \textbf{\textit{M}}_{\text{SKY}} 
\begin{bmatrix}
I_{\rm src} \\ Q_{\rm src} \\ U_{\rm src} \\ V_{\rm src} 
\end{bmatrix}.
\label{eq:main}
\end{equation}

\textbf{\textit{M}}$_\text{SKY}$ is a matrix accounting for the rotation in the linear polarization angle owing to the position angle ($\theta$) between the celestial sphere and the polarization feed probes of the receiver 
\begin{equation}
\textbf{\textit{M}}_{\text{SKY}} = 
\begin{bmatrix}
1 & 0 & 0 & 0 \\ 
0 & \cos 2\theta & \sin 2\theta & 0 \\ 
0 & -\sin 2\theta & \cos 2\theta & 0\\  
0 & 0 & 0 & 1 
\end{bmatrix}.
\label{eq:Msky}
\end{equation}

For an alt-az mounted telescope, $\theta$ is the parallactic angle. Unlike a typical radio telescope with a movable {dish}, the reflecting surface of FAST is mounted on a fixed frame on the ground while the 19-beam receiver is rotatable from $-80^\circ$ to $+80^\circ$ with respect to the equatorial North-South axis. Therefore, the $\theta$ for FAST is the rotation angle of the 19-beam receiver.

The \textbf{\textit{M}}$_\text{TOT}$ is composed of three matrices following the signal path in the receiving system. 
For a single linearly polarized feed horn, radiation first encounters the feed, then is affected by the imperfections of the feed, and finally proceeds through two independent amplifier chains, one for each orthogonal polarization. The total Mueller matrix of a receiving system is a product of $\textbf{\textit{M}}_\text{TOT} = \textbf{\textit{M}}_\text{A} \cdot \textbf{\textit{M}}_\text{IF} \cdot \textbf{\textit{M}}_\text{F} $.
The first matrix \textbf{\textit{M}}$_\text{F}$ represents the ability of one of the 19-beam feeds to convert the incoming pure linear polarization signal to any degree of elliptical polarization. The second matrix \textbf{\textit{M}}$_\text{IF}$ describes imperfections in the feed, specifically the production of nonorthogonal polarizations. The third matrix \textbf{\textit{M}}$_\text{A}$ represents the amplifier chain.

For a feed providing arbitrary elliptical polarization,
\begin{equation}
\textbf{\textit{M}}_{\text{F}} = 
\begin{bmatrix}
1 & 0 & 0 &0 \\
0 & \cos 2\alpha & \sin 2\alpha \cos \chi & \sin2\alpha\sin\chi \\
0 & -\sin2\alpha\cos\chi & \cos^2\alpha-\sin^2\alpha\cos2\chi & -\sin^2\alpha\sin2\chi \\
0 & -\sin2\alpha\sin\chi & -\sin^2\alpha\sin2\chi & \cos^2\alpha+\sin^2\alpha\cos2\chi 
\end{bmatrix},
\label{eq:Mf}
\end{equation}
where $\alpha$ is a measure of the voltage ratio of the polarization ellipse produced when the feed observes pure linear polarization, and $\chi$ is the phase angle of the two voltages specified by $\alpha$.
A perfect linear polarization feed has $\alpha = 0$ and $\chi = 0$. In reality, the output elliptical polarization can be misaligned with $X$ so that $\chi \neq 0$, but this is equivalent to having $\chi = 0$ and physically rotating the feed. Therefore, without loss of generality, we can take $\chi = 0$. 

For an imperfect feed generating output cross-correlated polarization from the input autocorrelated polarization,
\begin{equation}
\textbf{\textit{M}}_{\text{IF}} = 
\begin{bmatrix}
1 & 0 & 2 \epsilon\cos\phi & 2\epsilon\sin\phi \\
0 & 1 & 0 & 0 \\
2\epsilon\cos\phi & 0 & 1 & 0 \\
2\epsilon\sin\phi & 0 & 0 & 1 
\end{bmatrix},
\label{eq:Mif}
\end{equation}
where $\epsilon$ represents undesirable cross-coupling between the two input polarizations, and $\phi$ is the phase angle at which the voltage coupling $\epsilon$ occurs. 
This equation assumes that the coupling in a linear feed arises from the two probes being not quite orthogonal and the coupling in each has the same relative phase.
The first assumption is usually valid for a good feed with only a first order term of $\epsilon$, and the second assumption is valid with no loss in generality since the out-of-phase coupling case is included in \textbf{\textit{M}}$_\text{F}$.

The output polarizations of a feed next go through two independent amplifier chains of the two polarization paths with 
\begin{equation}
\textbf{\textit{M}}_{\text{A}} = 
\begin{bmatrix}
1 & \Delta G/2 & 0 & 0 \\
\Delta G/2 & 1 & 0 & 0 \\
0 & 0 & \cos\psi & -\sin\psi \\
0 & 0 & \sin\psi & \cos\psi 
\end{bmatrix}.
\end{equation}
$\Delta G$ is the error in relative intensity calibration of the two polarization paths. In practice, the amplifier gains are calibrated with a single noise-diode signal that is split and coupled into each of the independent polarization paths---in this way, the noise diode calibration signals injected into each path are correlated. If the ratio of the noise diode signal intensities is correct, $\Delta G = 0$. However, a small error in the relative noise diode intensities is expected owing to the instability of the noise diode itself. This allows us to assume $\Delta G \ll 0$ and only take the first order term of $\Delta G$. The relative phase between the two amplifier chains arising from the different path lengths of the two polarization paths can also be calibrated with the correlated noise diode signals. The remaining phase difference $\psi$ thus represents the phase difference between the incoming radiation from the sky and the noise diode.

Finally, 
\begin{equation}
\textbf{\textit{M}}_{\text{TOT}} = 
\begin{bmatrix}
1 & -2\epsilon\sin2\alpha\cos\phi+\frac{\Delta G}{2}\cos2\alpha &2\epsilon\cos2\alpha\cos\phi+\frac{\Delta G}{2}\sin2\alpha & 2\epsilon\sin\phi \\
\frac{\Delta G}{2} & -\Delta G\epsilon\sin2\alpha\cos\phi+\cos2\alpha & \Delta G\epsilon\cos2\alpha\cos\phi+\sin2\alpha & \Delta G\epsilon\sin\phi \\
2\epsilon\cos(\phi+\psi) & -\sin2\alpha\cos\psi & \cos2\alpha\cos\psi & -\sin\psi \\
2\epsilon\sin(\phi+\psi) & -\sin2\alpha\sin\psi & \cos2\alpha\sin\psi & \cos\psi 
\end{bmatrix}.
\label{eq:Mtot}
\end{equation}
In summary, the simplifications made in $\textbf{\textit{M}}_\text{TOT}$ are $\chi = 0$ and the ignorance of the second and higher order terms in $\epsilon$ and $\Delta G$. All orders in $\alpha$, $\phi$, and $\psi$ are retained in $\textbf{\textit{M}}_\text{TOT}$.  

\subsection{Solving the Mueller Matrix}
In practice, we cannot reliably measure the $\theta$ dependence of $I_{\rm obs}$ in Equation \ref{eq:main} because it is rendered inaccurate by small gain errors, either from the electronics or the zenith-angle gain dependence of the telescope. Thus, we use fractional correlator outputs and fractional source polarization:
\begin{equation}
\begin{bmatrix} 
1 \\ Q_{\rm obs}/I_{\rm obs} \\ U_{\rm obs}/I_{\rm obs} \\ V_{\rm obs}/I_{\rm obs} 
\end{bmatrix} 
=
\textbf{\textit{M}}_{\text{TOT}} \cdot \textbf{\textit{M}}_{\text{SKY}} 
\begin{bmatrix} 
1 \\ Q_{\rm src}/I_{\rm src} \\ U_{\rm src}/I_{\rm src} \\ V_{\rm src}/I_{\rm src} 
\end{bmatrix}.
\label{eq:main2}
\end{equation}
The signal-to-noise ratio in $I_{\rm obs}$ is much higher than those in $Q_{\rm obs}$, $U_{\rm obs}$, and $V_{\rm obs}$, and hence the errors in $Q_{\rm obs}/I_{\rm obs}$, $U_{\rm obs}/I_{\rm obs}$, and $V_{\rm obs}/I_{\rm obs}$ are dominated by the errors in $Q_{\rm obs}$, $U_{\rm obs}$, and $V_{\rm obs}$. Note that the division by $I_{\rm obs}$ produces errors in $Q_{\rm obs}/I_{\rm obs}$, $U_{\rm obs}/I_{\rm obs}$, and $V_{\rm obs}/I_{\rm obs}$, but these errors are second order because they are products of $\Delta G$ and/or $\epsilon$ that are already first order. Our treatment neglects all second order products, so we can neglect these errors. 
Further, the terms in the top row of Equation \ref{eq:Mtot} cause $I_{\rm obs}$ to not equal $I_{\rm src}$ for a polarized source. If one derives fractional polarization, then it will be in error by amounts comparable to $(\Delta G, \epsilon) \times (Q_{\rm src}, U_{\rm src}, V_{\rm src})$. For our sources with polarization percentages less than 10\%, these products are second order and therefore are of no concern.

\begin{figure}[ht!]
\centering
\includegraphics[width=0.35\textwidth]{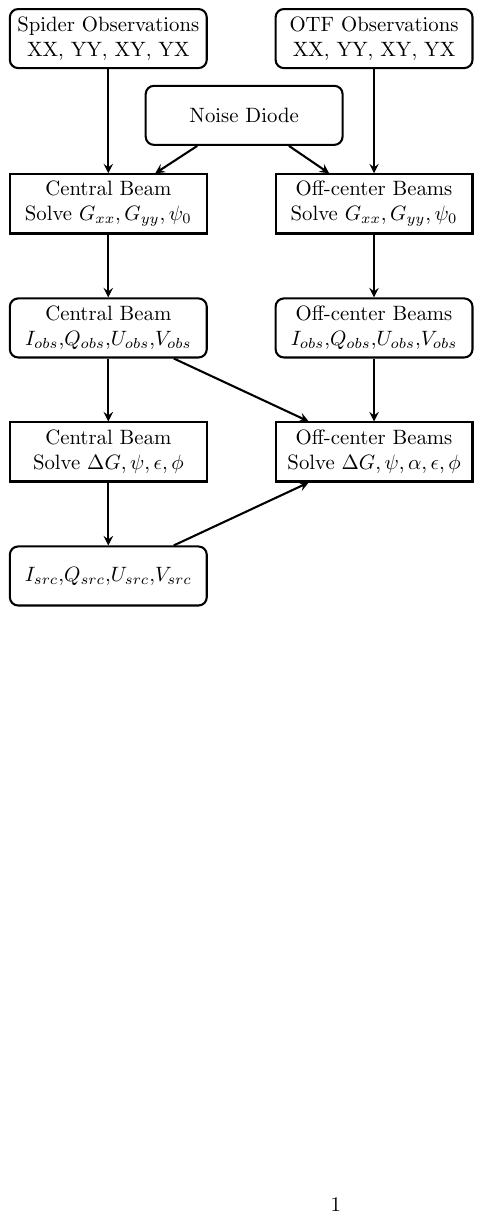}
\caption{Flow chart of the 19-beam polarization calibration sequence.
\label{fig:flowchart}}
\end{figure}

To calibrate the five parameters in Equation \ref{eq:Mtot}, one either observes a single polarization calibrator over a large range of $\theta$, or one observes several polarization calibrators with different polarization angles. Figure \ref{fig:flowchart} illustrates our polarization calibration sequence. We perform a spider observation toward a single polarization calibrator to calibrate the Mueller matrix of the central beam and OTF observations toward several polarization calibrators to calibrate the Mueller matrices of the 18 off-center beams. The injection of the correlated noise diode signal serves an initial calibration of the electronic gain and the relative phase between the two polarization paths.

\subsubsection{Noise Diode}
The feed’s response is modified by the electronics system, which introduces its own electronic gain and relative phase differences between the two paths. 
The electronic gain and relative phase can be effectively measured using the noise diode. 
Considering that the gain and phase may change rapidly with time, we use the modulated mode of the noise diode with a 2 seconds on/off switching period. 

Because the two polarization paths use independent amplifiers and other electronic devices, the $XX$ and $YY$ products have different electronic gains $G_{xx}$ and $G_{yy}$. The $G_{xx}$ and $G_{yy}$ are measured every 2 seconds using the diode deflection divided by the noise diode equivalent temperature. The $G_{xx}$ and $G_{yy}$ covert the self- and cross-products $(XX, YY, XY, YX)$ to units of kelvins.
Owing to the measurement errors in diode deflection and the time variability of noise diode equivalent temperature, the measurements of $G_{xx}$ and $G_{yy}$ are not perfect, and a relative gain between the two paths may remain. 
If the receiver is perfect and the relative gain differs, then the difference between the two paths is nonzero for an unpolarized source, making the source appear to be polarized.
The combination of the relative gain in the electronics path and the receiver gain difference between $XX$ and $YY$ is calibrated by $\Delta G$ in the Mueller matrix. 

The cable length difference due to thermal expansion and the phase delays produced by electronic devices such as amplifiers and bandpass filters generate a relative phase between the two polarization paths. For a perfect linearly polarized feed, if the relative phase of the two paths differs, then a purely linearly polarized source will appear to be elliptically polarized.
With the correlated noise diode, the relative phase $\psi_0$ is also measured every 2 seconds by $\psi_0 = \tan^{-1}(YX_{\rm on-off}/XY_{\rm on-off})$, where $YX_{\rm on-off}$ and $XY_{\rm on-off}$ are the diode deflections in $YX$ and $XY$, respectively. After the removal of $\psi_0$, the remaining instrumental phase in the $XY$ and $YX$ products is dominated by the receiver phase $\psi$, which is calibrated in the Mueller matrix.

\subsubsection{Spider Observations}\label{sec:spider}
\begin{figure}[h!]
\plotone{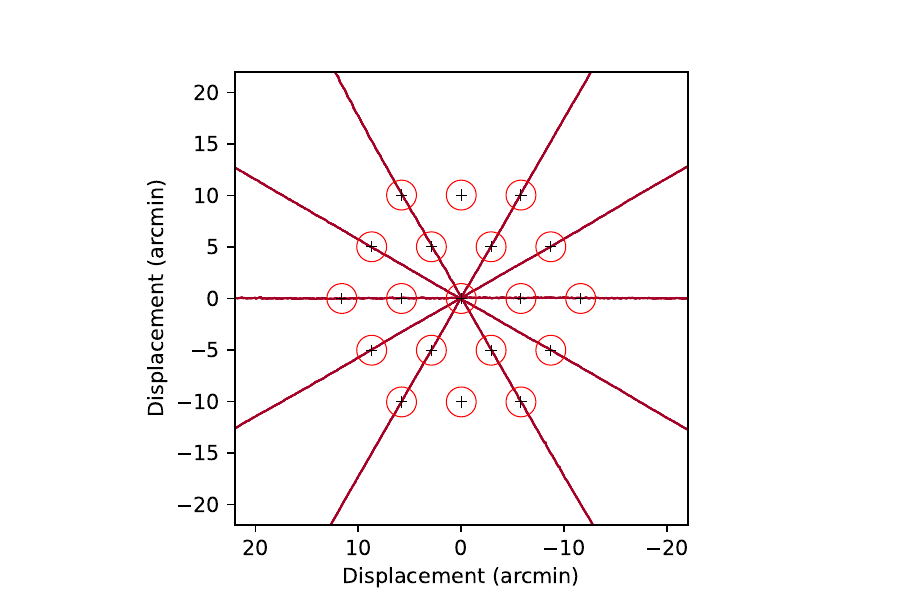}
\caption{Real-time tracks of the spider observation toward 3C286 on 2020/09/12. The tracks were drifted along the equatorial East-West axis.
{For example, the beams M08, M02, M01, M05, and M14 are aligned from East to West as Figure \ref{fig:19beam} when receiver rotation angle $\theta = 0^{\circ}$, but the beams M18, M07, M01, M04, and M12 are aligned from East to West when $\theta = 60^{\circ}$.
Here we show the relative trajectories of the tracks with respect to the receiver center with $\theta$ at $-60^{\circ}, -30^{\circ}, 0^{\circ}, 30^{\circ},$ and 60$^{\circ}$.} 
\label{fig:sp_track}}
\end{figure}

A spider observation consists of several scans of a polarized source drifted over a wide range of position angles. It is the standard technique for the polarization calibration of the single-pixel receivers of Arecibo \citep{2001aHeiles} and the GBT \citep[GBT Commissioning Memo\footnote{https://library.nrao.edu/gbtcm.shtml}: 23;][]{2016Liao,2023Fallon}. Figure \ref{fig:sp_track} shows an example of the real-time tracks of a spider observation. We perform a spider observation toward a linearly polarized source with the receiver rotation angle $\theta$ at $-60^{\circ}, -30^{\circ}, 0^{\circ}, 30^{\circ},$ and 60$^{\circ}$. For example, when the receiver is rotated by 60$^{\circ}$, beams M01, M04, M07, M12, and M18 are aligned along the equatorial East-West axis. A full spider procedure takes 58 minutes with the middle of the time interval allocated close to the transit of the polarization calibrator.

Given the assumption of $\chi = 0$ in \textbf{\textit{M}}$_\text{F}$, Equation \ref{eq:Mf} becomes a rotation matrix of angle {2$\alpha$}.
{Since Equation \ref{eq:Msky} and Equation \ref{eq:Mf} with $\chi = 0$ are both rotation matrices, the combination of the two matrices is a rotation matrix of  angle $2\alpha + 2\theta$. The $\alpha$ and $\theta$ thus are coupled when solving the Mueller matrix.} 
Since $\theta$ may have a small systematic error in the mechanical control of the rotation, $\alpha$ is hard to be accurately determined. We thus fix $\alpha = 0$ for the central beam. The spider observation can simultaneously obtain not only the Mueller matrix parameters $(\Delta G, \phi, \epsilon, \psi)$ but also the Stokes parameters of the source $(Q_{\rm src}/I_{\rm src}, U_{\rm src}/I_{\rm src}, V_{\rm src}/I_{\rm src})$. Although a spider observation provides accurate calibration, carrying out spider observations for all the 19 beams is inefficient. We thus only perform spider observation of the central beam and use the accurately determined $Q_{\rm src}/I_{\rm src}$, $U_{\rm src}/I_{\rm src}$, and $V_{\rm src}/I_{\rm src}$ as known parameters when calibrating the 18 off-center beams with OTF observations.

\subsubsection{OTF Observation}
\begin{figure}[h!]
\plotone{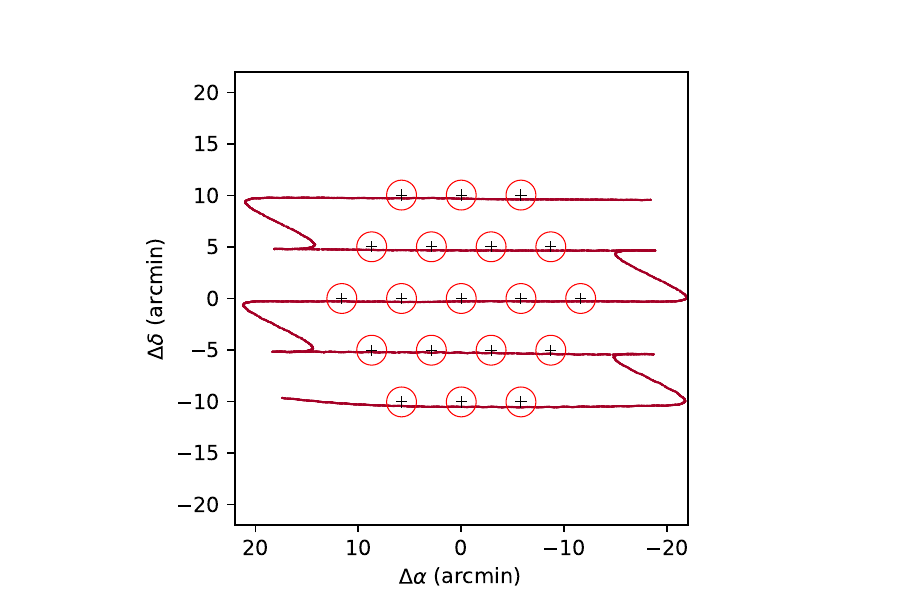}
\caption{A real-time track of the OTF observation toward 3C138 on 2021/10/07.  
\label{fig:otf_track}}
\end{figure}

Figure \ref{fig:otf_track} shows an example of the real-time track of an OTF observation. Our regular setup of the scan gap between the rows is 5.0 arcmin to pass through the central positions of the beams, but a scan gap of 1.0 arcmin is used for the J0854+2006 observations to map the beam structures (see {Paper II}). Depending of the ZA of the calibrator, the OTF observation with 5.0 arcmin gaps takes about 13 minutes, and an OTF observation with 1.0 arcmin gaps takes about 2 hours. The receiver rotation angle $\theta$ is fixed at 0$^{\circ}$ during an OTF observation. The $I_{\rm obs}$, $Q_{\rm obs}$, $U_{\rm obs}$, and $V_{\rm obs}$ data from OTF observations and spider observations are used together to solve the Muller matrices of the 18 off-center beams. With the polarization angles of the calibrators obtained from the central beam calibration, we can keep $\alpha$ a free parameter when solving the Mueller matrix, and the $\alpha$ represents the total effect of voltage ratio and mechanical rotation error of the beam with respect to that of the central beam. 

\section{Results} \label{sec:res}

\subsection{Central Beam}

\begin{figure}[h!]
\epsscale{.7}
\plotone{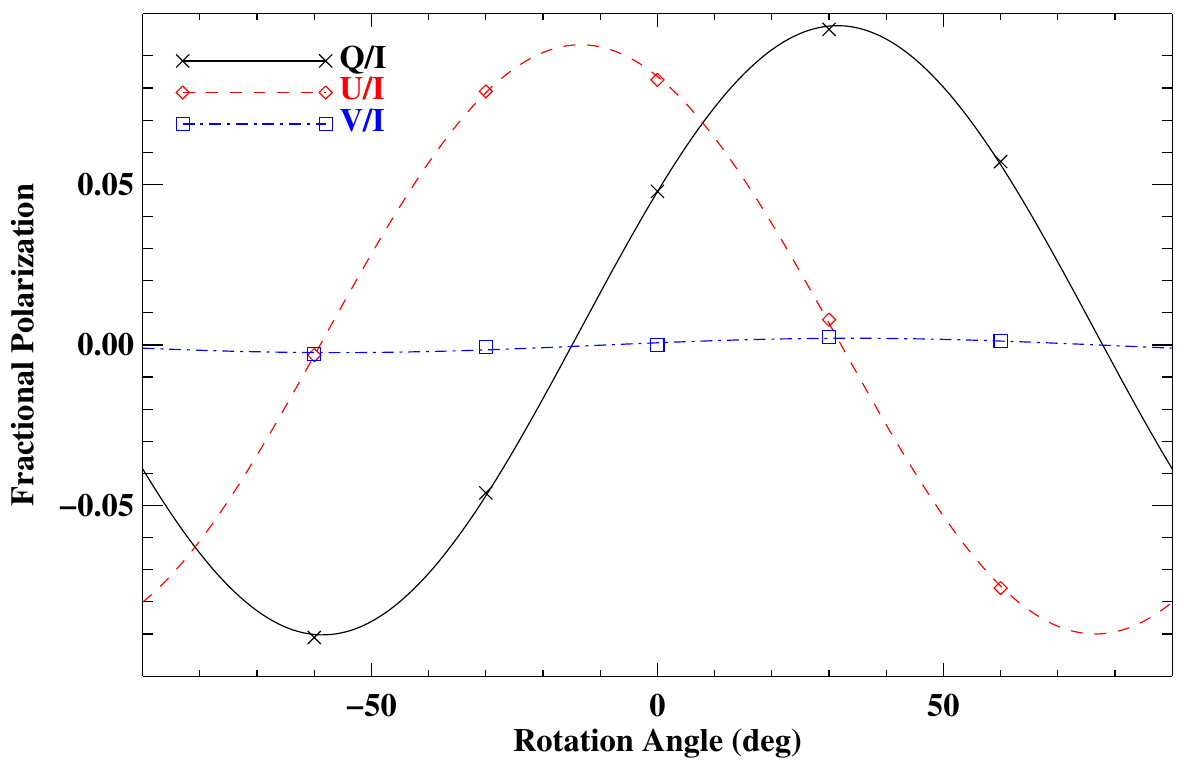}
\caption{The fractional Stokes parameters for 3C286 versus rotation angle $\theta$ on 2023/03/11. The {black crosses, red diamonds, and blue squares} show the $Q_{\rm obs}/I_{\rm obs}$, $U_{\rm obs}/I_{\rm obs}$, and $V_{\rm obs}/I_{\rm obs}$ data, respectively. The curves show the fittings for $Q_{\rm src}/I_{\rm src}$, $U_{\rm src}/I_{\rm src}$, and $V_{\rm src}/I_{\rm src}$ in {black, red,} and blue colors.\label{fig:RHSTK}}
\end{figure}

\begin{figure}[h!]
\plotone{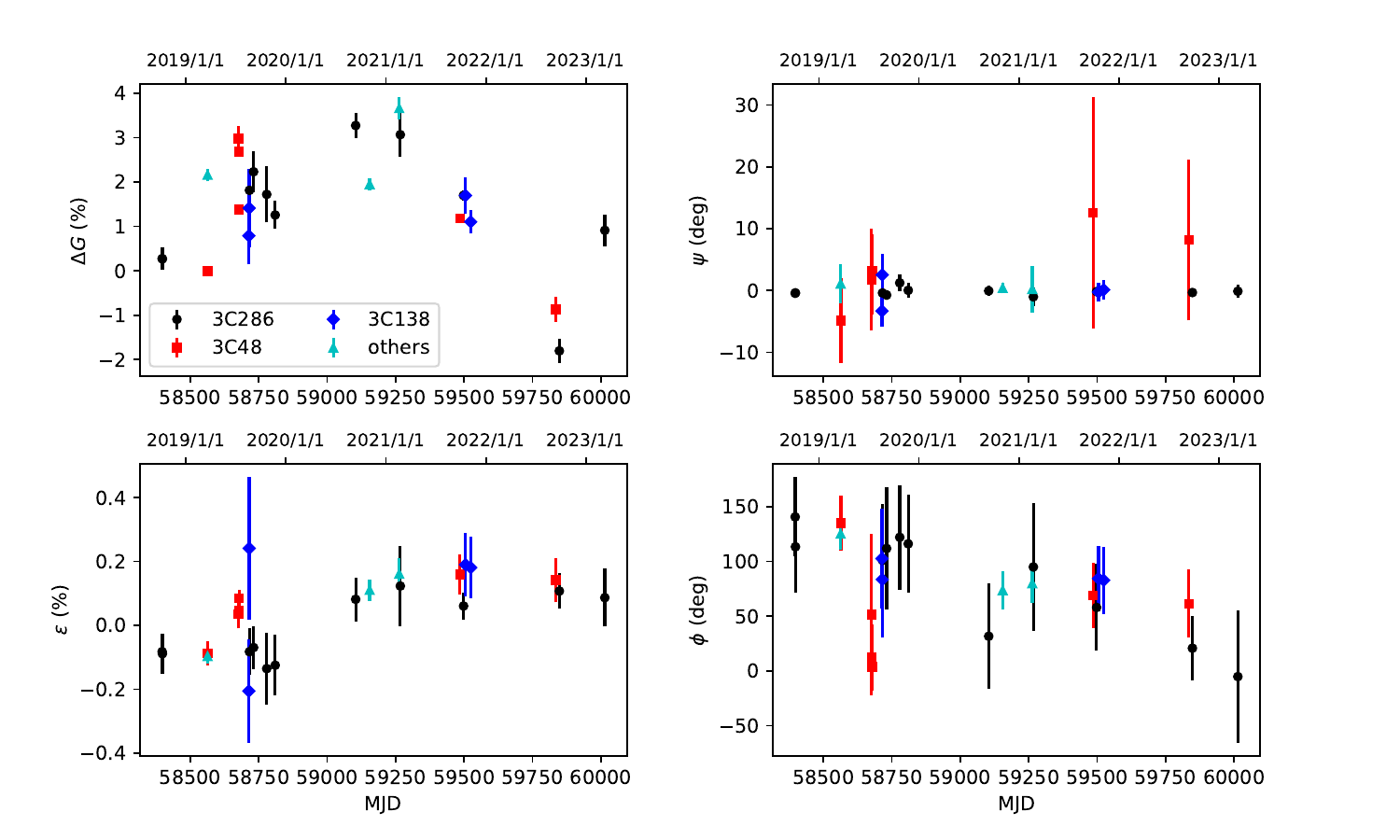}
\caption{Parameters $(\Delta G, \psi, \epsilon, \phi)$ of the central beam versus observation dates in Modified Julian Day (MJD). The results of 3C286, 3C48, and 3C138 are plotted as {black circles, red squares, and blue diamonds}, respectively. The results of 3C380, J2202+4216, and 3C273 are shown as {cyan triangles}.\label{fig:mm_paras_center}}
\end{figure}

Figure \ref{fig:RHSTK} shows an example of the observed and fitted fractional Stokes parameters as a function of rotation angle $\theta$ from a spider observation. The fitting of the Mueller matrix (Equation \ref{eq:Mtot}) with assumption of $\alpha = 0$ gives a set of parameters $(\Delta G, \psi, \epsilon, \phi)$.  
Figure \ref{fig:mm_paras_center} shows the Mueller matrix parameters obtained from the spider observations from October 2018 to March 2023.
The values of $\Delta G$ have a range from $-2$\% to 4\%. The typical value of $\psi$ is between $-5^{\circ}$ and $5^{\circ}$. Owing to the weak polarization percentage of 3C48, the absolute values and errors of the $\psi$ obtained from 3C48 are larger than those from other sources. 
The typical value of $\epsilon$ is between $-0.2$\% and 0.2\%, except the 3C138 data in July 2019 at ZAs larger than $26.4^{\circ}$. Because $\epsilon$ and $\phi$ are coupled in terms of $\epsilon \sin \phi$ and $\epsilon \cos \phi$ in Equation \ref{eq:Mif}, the small range of $\epsilon$ lead to a large range of $\phi$. 
The $\phi$ hence shows a range from $-20^{\circ}$ to $150^{\circ}$ with errors of about $30^{\circ}$, much larger than the ranges and errors of other parameters. Some of the Mueller matrix parameters appear to have correlations with time. There is a strong increasing trend of $\epsilon$ from $-1$\% in late 2019 to 2\% in late 2021 and an overall decreasing trend of $\phi$ from $150^{\circ}$ to $-20^{\circ}$ in the whole period of the data. The $\Delta G$ appears to increase from late 2018 to late 2020 and decrease from early 2021 to late 2022. A detailed analysis of the autocorrelation of $\Delta G$ is presented in Section \ref{sec:app}.

\begin{figure}[h!]
\plotone{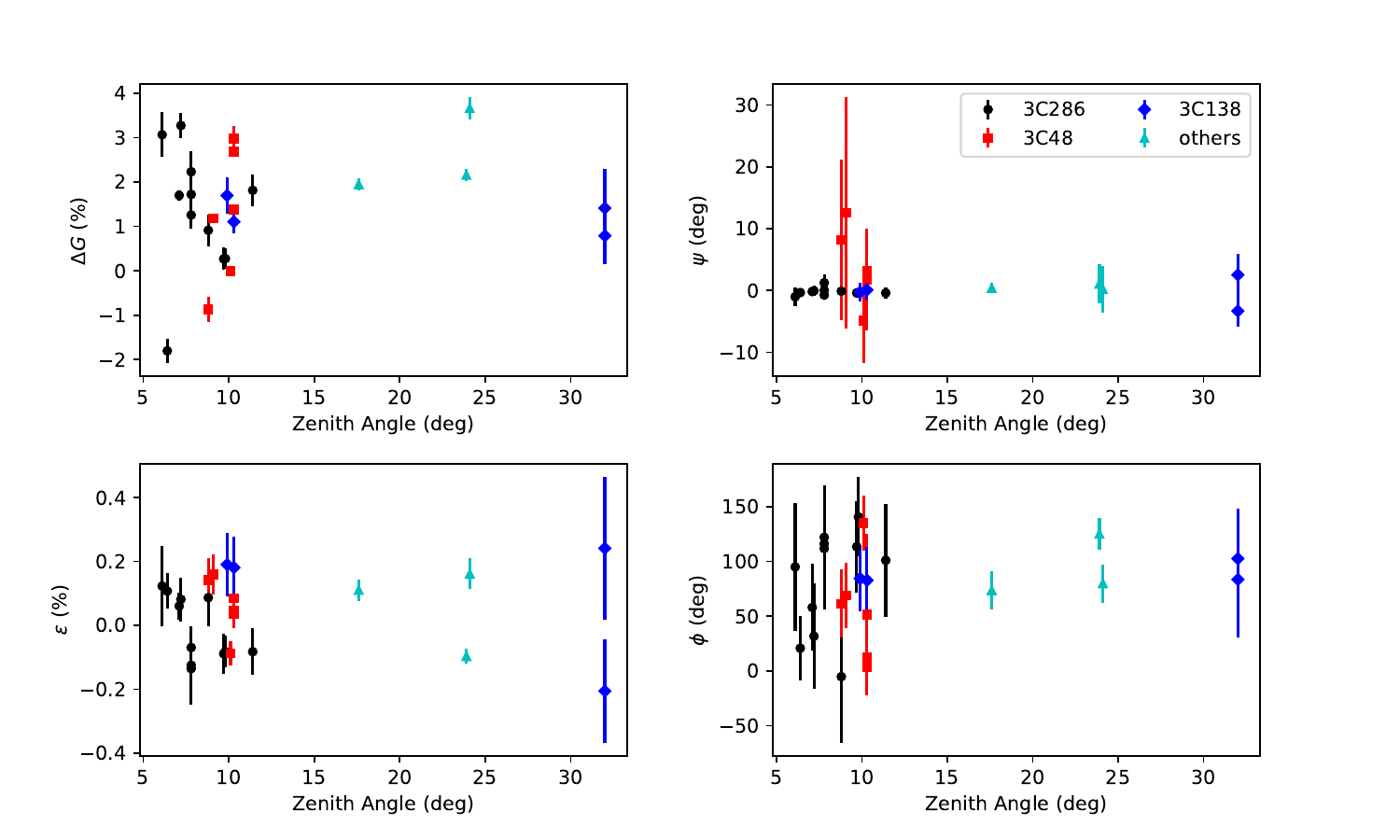}
\caption{The same data as Figure \ref{fig:mm_paras_center} but versus ZA.\label{fig:mm_paras_za}}
\end{figure}

FAST uses different portions of the {reflector surface} toward different sky positions, and thus the different declinations of 3C48, 3C138, and 3C286 at $+33.2^{\circ}$, $+30.5^{\circ}$, and $+16.6^{\circ}$ help to diagnose the dependence of Mueller matrix parameters on the {reflector surface}. If the Mueller matrix parameters strongly depend on the {reflector surface}, the 3C48, 3C138, and 3C286 data in Figure \ref{fig:mm_paras_center} taken on nearby observation dates should have large deviations from each other. The consistent values of the Mueller matrix parameters on nearby dates suggest that the Mueller matrix parameters of the central beam at ZA $<26.4^{\circ}$ do not have a strong dependence on the {reflector surface}. Figure \ref{fig:mm_paras_za} shows the Mueller matrix parameters as a function of ZA. For ZA $< 26.4^{\circ}$, the data appear to be scattered without a strong dependence on ZA. The 3C138 data at large ZAs show larger errors than other data, and the $\psi$ and $\epsilon$ of the 3C138 data at large ZAs also appear more scattered than other data. The Mueller matrix parameters at large ZAs need to be further measured and analyzed. 

\begin{figure}[h!]
\plotone{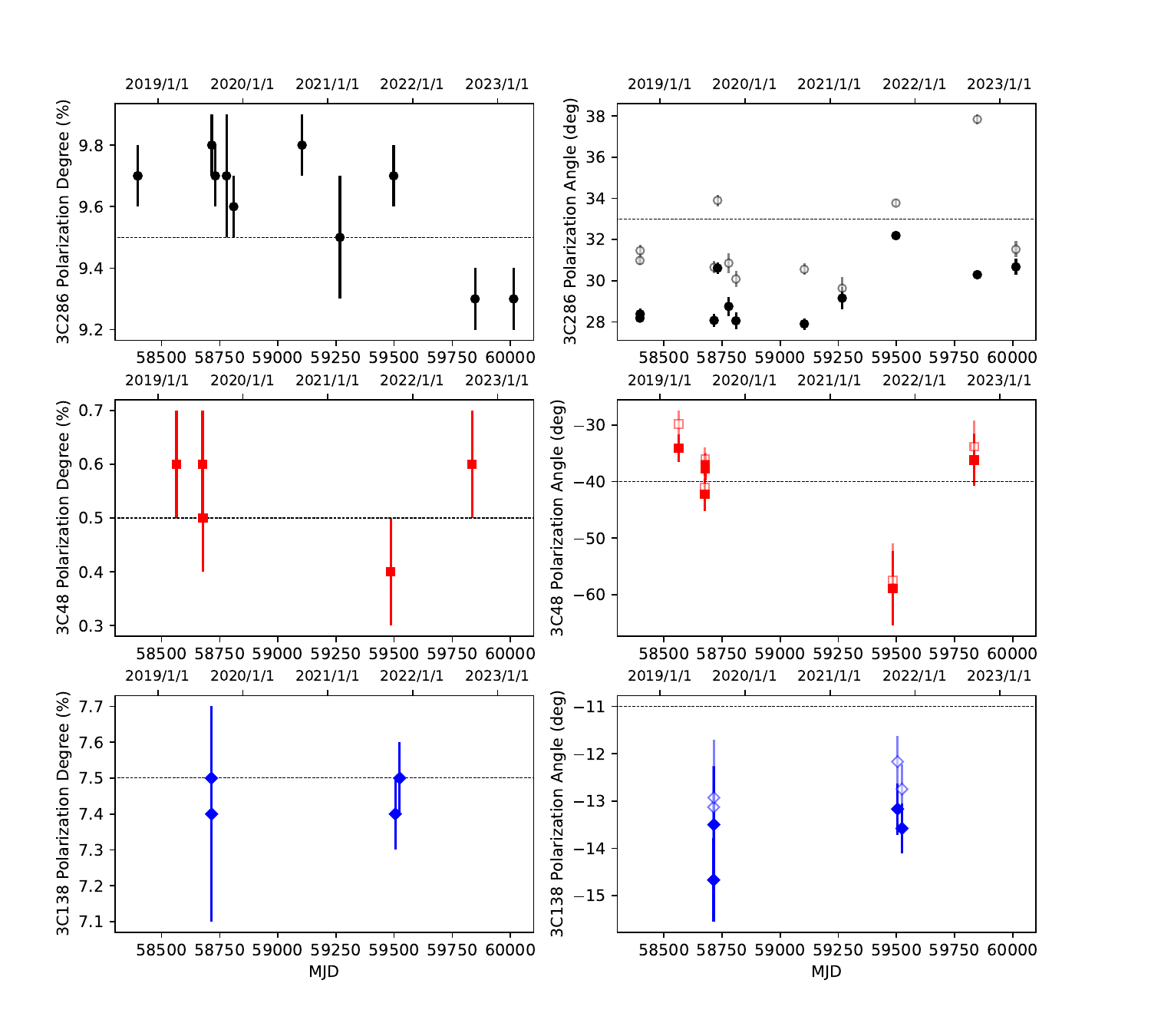}
\caption{Calibrated polarization percentages and polarization angles of 3C286, 3C48, and 3C138 in the same color scheme as that in Figures \ref{fig:mm_paras_center} and \ref{fig:mm_paras_za}. The polarization angles before and after the correction of the ionospheric Faraday rotation are denoted with open and filled {data points}, respectively. The dashed lines mark the VLA results at 1450 MHz in December 2010 \citep{2013PB}. \label{fig:3c}}
\end{figure}

Figure \ref{fig:3c} shows the polarization percentages and polarization angles of 3C286, 3C48, and 3C138 derived from the calibrated $Q_{\rm src}/I_{\rm src}$, $U_{\rm src}/I_{\rm src}$, and $V_{\rm src}/I_{\rm src}$ data. We also correct the ionospheric Faraday rotation in the polarization angles using the weighted average of dozens of global ionospheric maps from different analysis centers \citep{2024Meng}.
For the data at ZA $<26.4^{\circ}$, the uncertainties in polarization percentages are between 0.1\% to 0.2\%, and the uncertainties in polarization angles are about 0.5$^{\circ}$ for 3C286 and 3C138 and about 3$^{\circ}$ for 3C48. The large uncertainties in 3C48 polarization angles are mainly the result of the large errors in $\psi$.
The uncertainties in the polarization percentages and polarization angles of 3C138 data at ZA $> 26.4^{\circ}$ are significantly larger than the 3C138 data at ZA $<26.4^{\circ}$. 
The polarization percentages of our results are consistent with the VLA results in \citet{2013PB} with a difference of about 0.2\%. The polarization angles of 3C48 are roughly consistent between the FAST and VLA results, but probably owing to the time variability of 3C286 and 3C138 polarization angles, the 3C286 and 3C138 polarization angles of the FAST data from 2018 to 2023 are different by about 3$^{\circ}$ from the 2010 VLA data. 

\subsection{Off-center Beams}
\begin{figure}[h!]
\epsscale{1.2}
\plotone{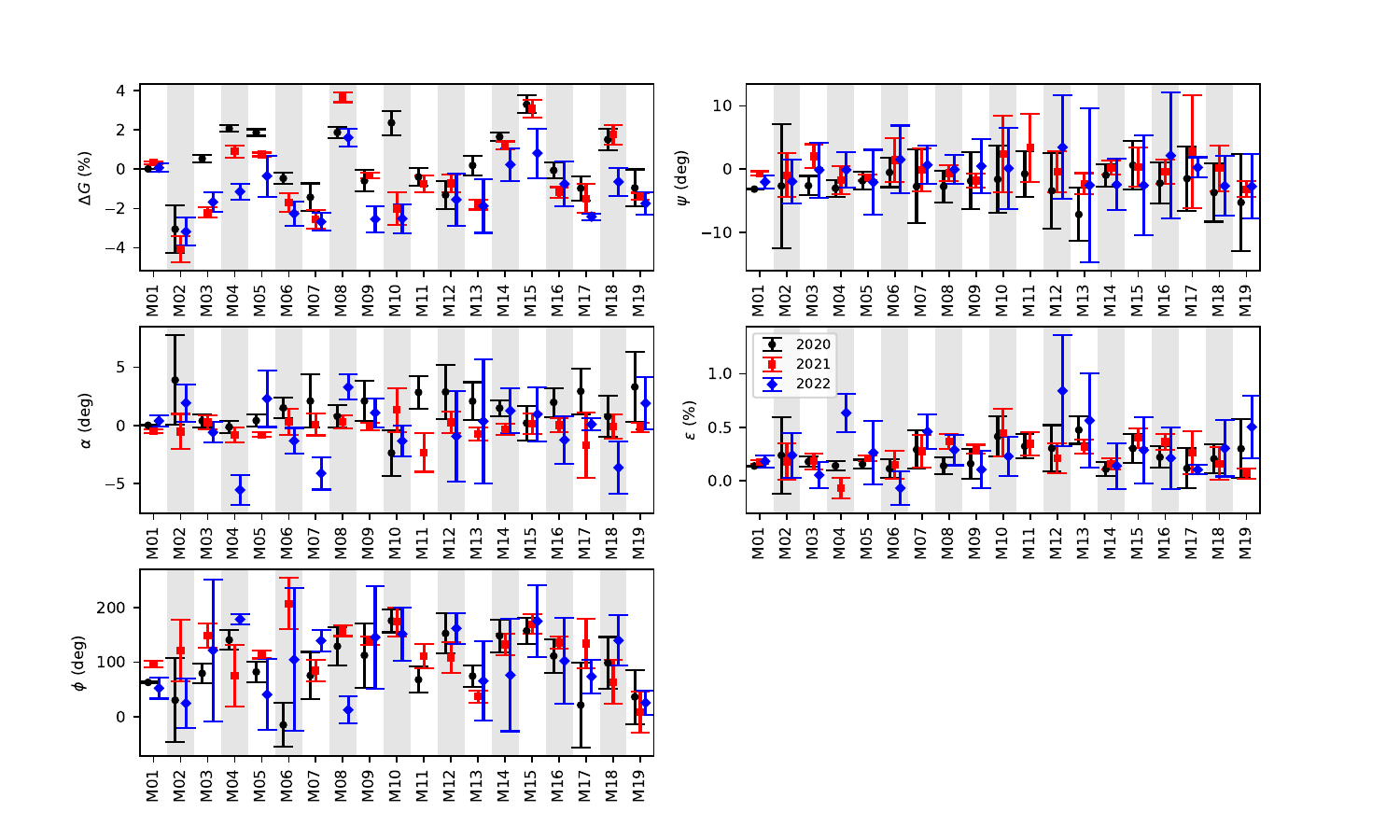}
\caption{Parameters $(\Delta G, \psi, \alpha, \epsilon, \phi)$ of the 19 beams. The parameters calibrated in 2020, 2021, and 2022 are shown in black circles, red squares, and blue diamonds, respectively.\label{fig:mm_paras_18}}
\end{figure}

Figure \ref{fig:mm_paras_18} shows the Mueller matrix parameters $(\Delta G, \psi, \alpha, \epsilon, \phi)$ of the 19 beams from 2020 to 2022.
The parameters in 2020 are obtained from the combination of the OTF observations in September and the 3C286 spider observation on 2020/09/12.
The parameters in 2021 are obtained from the combination of the OTF observations in October and the spider observations toward 3C48, 3C286, and 3C138 on 2021/09/28, 2021/10/11, and 2021/10/18. 
The parameters in 2022 are obtained from the combination of the OTF observations in September and the 3C286 spider observations on 2022/09/26. 
Owing to an interruption at the end of the 3C48 OTF observation on 2022/09/19, the M10, M11, and M12 data were missed. Because the M11 data in 2022 were only available from the 3C138 and J0854+2006 OTF observations, we cannot obtain accurate fitting of the Mueller matrix parameters of M11. Hence, the M11 2022 data in Figure \ref{fig:mm_paras_18} are missing. 

The $Q_{\rm src}/I_{\rm src}$, $U_{\rm src}/I_{\rm src}$, and $V_{\rm src}/I_{\rm src}$ of the OTF sources are obtained by applying the central-beam Mueller matrix on the central-beam $Q_{\rm obs}/I_{\rm obs}$, $U_{\rm obs}/I_{\rm obs}$, and $V_{\rm obs}/I_{\rm obs}$ data, and then the $Q_{\rm src}/I_{\rm src}$, $U_{\rm src}/I_{\rm src}$, and $V_{\rm src}/I_{\rm src}$ are used to derived the Mueller matrix parameters of the off-center beams using Equation \ref{eq:main2}.
Similar to a self-calibration of the central beam, the errors in the central-beam Mueller matrix parameters therefore are significantly smaller than the those of the off-center beams. The differences in Mueller matrix parameters between the central beam and off-center beams in Figure \ref{fig:mm_paras_18} also demonstrate the different accuracy of polarization calibration between spider observations and OTF observations. 

The Mueller matrix parameters in 2021 generally have smaller errors than those in 2020 and 2022 for two reasons. First, the 2021 parameters are fitted with 3 spider observations and 5 OTF observations, more than the numbers of observations in 2020 and 2022. Second, the track of a spider observation passes the center of the beams closer than the track of an OTF observation (Figures \ref{fig:sp_track} and \ref{fig:otf_track}), hence the observed Stokes parameters in a spider observation are less contaminated by the off-axis structures in Stokes $I, Q, U, V$ beams (see the beam patterns in Paper II) than the observed Stokes parameters in an OTF observation. The three spider observations used to derive the 2021 parameters thus help to produce small errors.
For the 2021 M11 and M17 data where spider observations were not available, the errors are generally larger than other beams in 2021 and comparable to the errors in 2020 and 2022.

Some parameters of the off-center beams are consistent from 2020 to 2022 with the distribution of the three-year values roughly within their errors, such as the $\psi$ of M02, M06, M07, M10, M15, the $\alpha$ of M15, the $\epsilon$ of M02, M14, M18, and the $\phi$ of M15 and M19. 
Similar to the $\phi$ in Figure \ref{fig:mm_paras_center} which shows an overall trend in time, some parameters of the off-center beams show increasing or decreasing trends from 2020 to 2022, such as the $\Delta G$ of M04, M05, M14 and the $\alpha$ of M07 and M18. 
In general, the Mueller matrix parameters of the off-center beams are not as accurate as those of the central beam to study their long-term performance in detail. 

\begin{figure}[h!]
\epsscale{.7}
\plotone{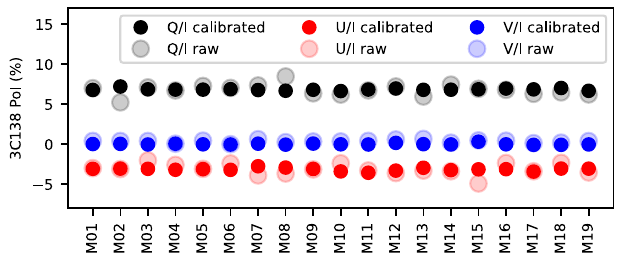}
\caption{The 19-beam $Q_{\rm obs}/I_{\rm obs}$, $U_{\rm obs}/I_{\rm obs}$, and $V_{\rm obs}/I_{\rm obs}$ data of {the} 3C138 on 2021/10/06, shown in black, red, and blue colors, respectively. The data before polarization calibration are shown in fainter colors than the data after polarization calibration.\label{fig:mm_3c138}}
\end{figure}

Figure \ref{fig:mm_3c138} shows the fractional Stokes parameters of 3C138 OTF observation before and after polarization calibration. A perfect polarization calibration should produce identical fractional Stokes parameters from M01 to M19. The 3C138 19-beam data after polarization calibration are more identical than the data before polarization calibration, indicating that the polarization calibration works properly.  

\section{Analysis and Application} \label{sec:app}
\subsection{Noise Diode Fluctuation}
$\Delta G$ is unique among the five Mueller matrix parameters in that its exact value depends only on the equivalent temperatures of the noise diode. A nonzero value of $\Delta G$ is usually not an intrinsic property of the receiver, but rather it depends on the equivalent temperatures and their frequency dependence.
We used the measurements of noise diode equivalent temperatures from the observatory in August 2018, January 2019, May 2020, and October 2020.
The measured values usually have nontrivial uncertainties of about a few percent. Moreover, the noise diode intensities vary unpredictably with time and frequency.
Our choices of the values for these equivalent temperatures have inevitable errors in the few-percent range. During the data reduction process, the errors in equivalent temperatures transfer to the nonzero values in $\Delta G$.

\begin{figure}[h!]
\plotone{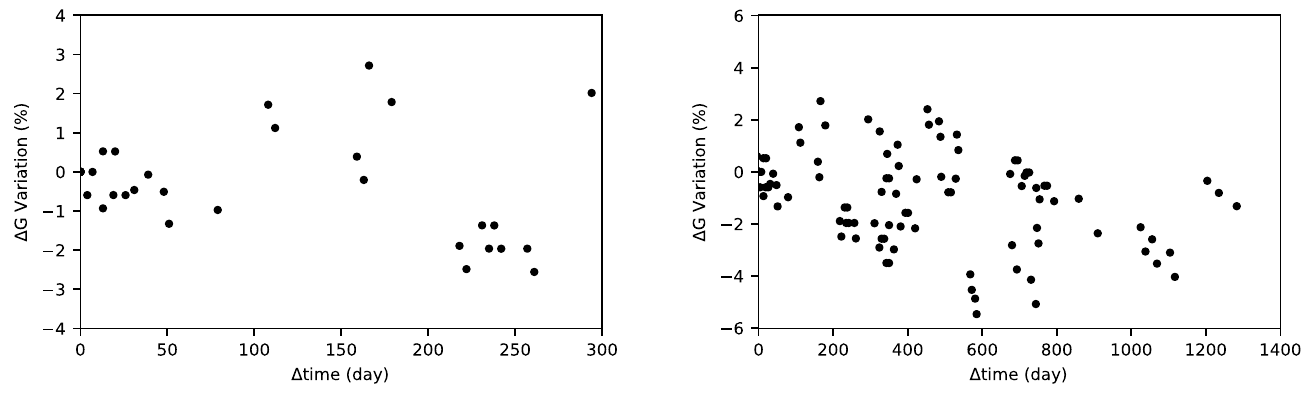}
\caption{Autocorrelation between any two data points of $\Delta G$ in Figure \ref{fig:mm_paras_center} by taking their difference in $y$-axis as variation and their difference in $x$-axis as time separation $\Delta$time. Left: the correlations with $\Delta$time shorter than 300 days. Right: the correlations of whole data sets. \label{fig:deltaG}}
\end{figure}

To study the fluctuation in noise diode, we derive the autocorrelation of the $\Delta G$ data in Figure \ref{fig:mm_paras_center}.
Figure \ref{fig:deltaG} shows the autocorrelation of $\Delta G$. 
The autocorrelation shows a broadening trend from $\sim$1\% variation at $\Delta$time close to 0 days to a variation about $\pm3\%$ at $\Delta$time of 300 days.
The autocorrelation also shows a long-term decreasing trend to a variation of $-4\%$ at $\Delta$time longer than $\sim$500 days.
In the equivalent temperature measurements, the fluctuation in noise diode is smaller than 1\% over 3.5 hours \citep{2020Jiang}.
Among the four equivalent temperature measurements from August 2018 to October 2020, the variation of the equivalent temperatures is about 2\% in a year. Our analysis of the $\Delta G$ autocorrelation is in agreement with the level of noise diode fluctuation, indicating that the $\Delta G$ is dominated by the noise diode rather than the receiver.
 
\subsection{Dependence of Mueller Matrix on the {Reflector Surface}}
\begin{figure}[h!]
\centering
\includegraphics[width=0.6\textwidth]{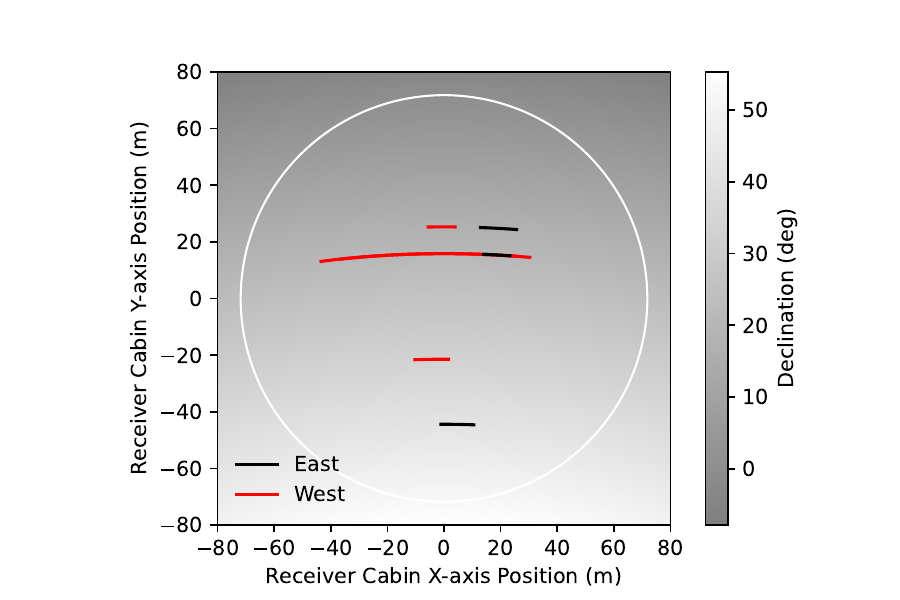}
\caption{Tracks of the receiver in geographic coordinates relative to the {reflector surface} center of FAST. The white circle represents the limit of full illumination at ZA $= 26.4^{\circ}$.\label{fig:drifts}}
\end{figure}

\begin{figure}[h!]
\epsscale{1.2}
\plotone{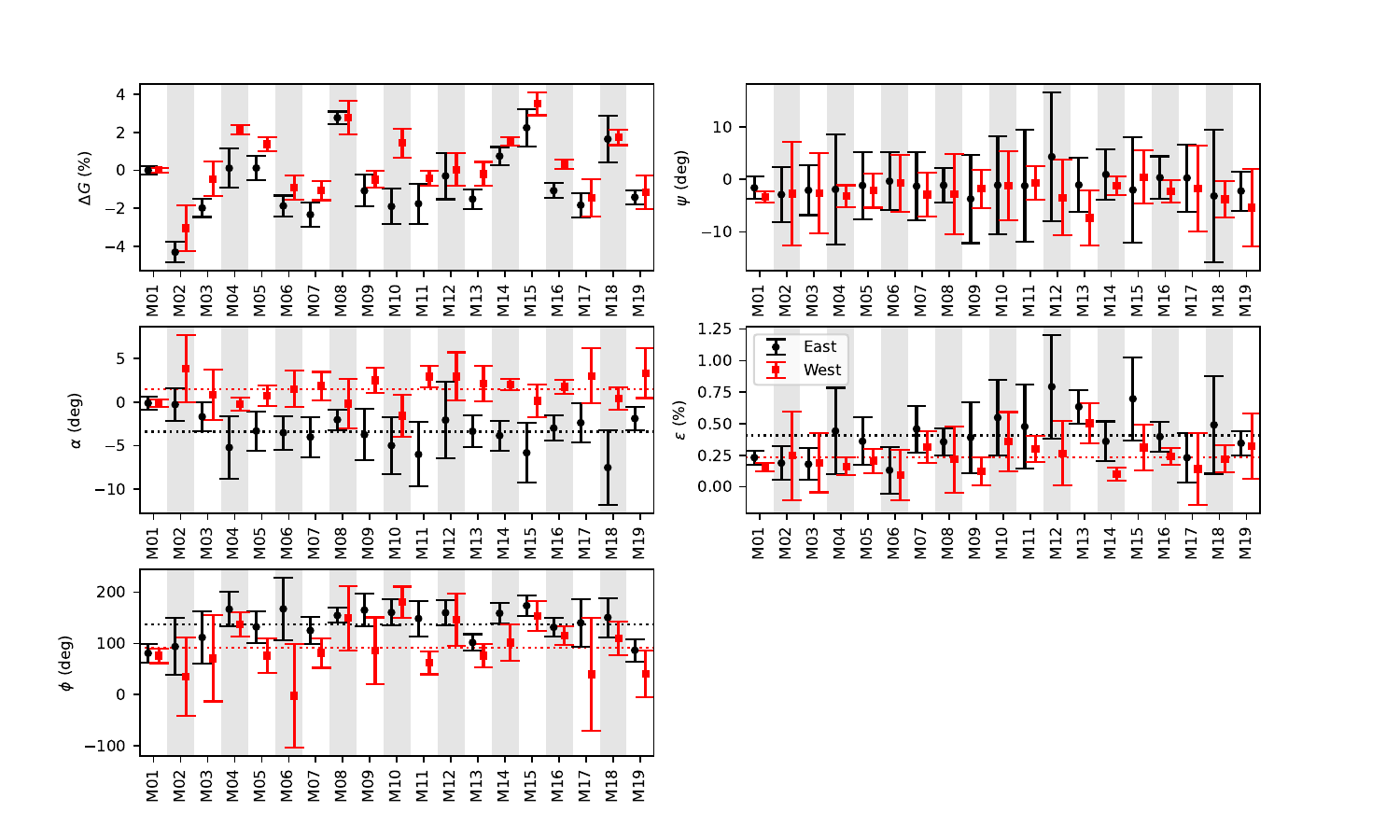}
\caption{Parameters $(\Delta G, \psi, \alpha, \epsilon, \phi)$ for the 19 beams on the eastern (black circles) and western (red squares) sides of the FAST {reflector surface}. The averaged values of the 19 beams of $\alpha$, $\epsilon$, and $\phi$ are shown as dotted lines.\label{fig:mm_paras_ew}}
\end{figure}

As radiation travels in a telescope, the polarization is sequentially affected by the telescope structure, {reflector surface}, receiver, and electronics system. 
The calibration of $\Delta G$ and $\psi_0$ can eliminate the relative gain and phase induced by the electronics system. 
Without a strong dependence on ZA, the parameters $(\psi, \epsilon, \phi)$ of the central beam at ZA $<$ 26.4$^{\circ}$ appear to be dominated by the receiver.
However, we notice that the Mueller matrix parameters of the off-center beams have a dependence on the {reflector surface}.
We compare the Mueller matrix parameters derived from the observations of 3C84 on 2021/10/05, 3C138 on 2021/10/06, and J0854+2006 on 2022/09/20 that use the eastern side of the {reflector surface} and the observations of J0854+2006 on 2020/09/10 and 2020/09/11, 3C138 on 2020/09/11, and 3C48 on 2021/10/10 that use the western side of the {reflector surface}.
Figure \ref{fig:drifts} shows the tracks of those observations on the FAST {reflector surface}, and Figure \ref{fig:mm_paras_ew} shows the Mueller matrix parameters derived from those observations. The parameters $(\alpha, \epsilon, \phi)$ of the observations using the eastern side of the {reflector surface} are different to those using the western side. The 19-beam averaged $\alpha$ using the eastern and western sides of the {reflector surface} are different by $\sim$5$^{\circ}$, $\epsilon$ differs by $\sim$0.2\%, and $\phi$ differs by $\sim$50$^{\circ}$.
Although these differences are not very significant compared to the errors of the parameters, the dependence of the Mueller matrix on the {reflector surface} is very likely to be enhanced at ZA $>26.4^{\circ}$, particularly because of the backward illumination mode.

\subsection{Mean 19-beam Mueller Matrices from 2020 to 2022}
Table \ref{tab:mm} lists the average parameters of the 19-beam Mueller matrices using the data from 2020 to 2022 in Figure \ref{fig:mm_paras_18}.
The $(\overline{\Delta G},\overline{\psi},\overline{\alpha},\overline{\epsilon},\overline{\phi})$ are the means weighted by the inverse square of the errors, and $(\sigma_{\overline{\Delta G}}, \sigma_{\overline{\psi}},\sigma_{\overline{\alpha}},\sigma_{\overline{\epsilon}},\sigma_{\overline{\phi}})$ are the weighted standard deviations. By taking the inverse matrices of $\textbf{\textit{M}}_{\text{TOT}}$ and $\textbf{\textit{M}}_{\text{SKY}}$, one should be able to obtain intrinsic Stokes parameters from observed Stokes parameters with Equation \ref{eq:main}. Since the signal in two polarization paths are split to the pulsar backend and the spectral-line backend right before the backends, the parameters here obtained using the spectral-line backend should be applicable to pulsar data. However, there are several caveats in using the Mueller matrix parameters.
First, the relative phase $\psi_0$ needs to be calibrated before applying the Mueller matrix.
Second, the parameters are only applicable to observations at ZA $<$ 26.4$^{\circ}$ between 2020 and 2022.
Last, the upper limits of $\sigma_{\overline{\Delta G}}$ and $\sigma_{\overline{\epsilon}}$ of the beams are about 2\% and 0.3\%, implying that after applying the Mueller matrix parameters in Table \ref{tab:mm}, the uncertainty of the leakage from Stokes $I$ to Stokes $Q$ can be as large as 2\% in amplitude, and the uncertainty of the leakage from Stokes $I$ to Stokes $U$ and $V$ can be as large as 0.3\% in amplitude.
Since these uncertainty levels are the major sources of errors in polarization measurements after applying Table \ref{tab:mm}, the errors are about 2\% in fractional linear polarization and 0.3\% in fractional circular polarization.
That is, considering a signal-to-noise ratio of five, only the on-axis fractional linear polarization measurements $\gtrsim$ 10\% and on-axis fractional circular polarization measurements $\gtrsim$ 1.5\% can be considered high confident detections. For on-axis polarization measurements below these levels, polarization calibration using spider observations which can calibrate $\Delta G$ more accurately than 0.2\% and $\epsilon$ more accurately than 0.1\% must be performed close to the dates of the scientific observations.

In this work, we state the amplitude of the changes in a number of the Mueller matrix parameters. These parameters should be tied to instrumental properties of the receiver. From 2018 to 2023, the receiver and the noise diode had been always installed in the receiver cabin with no significant changes. The only major maintenance of FAST was a replacement of the cables connecting the receiver cabin and six support towers between 2020/06/10 and 2020/07/28. Although Figure \ref{fig:mm_paras_center} shows a noticeable turning point of $\Delta G$ from an increasing trend before middle 2020 to a decreasing trend after middle 2020, the other Mueller matrix parameters do not show strong variations before and after the cable replacement.
At this moment, we do not have enough data to study the origins of the time variations of the Mueller matrix parameters. We will try to address this topic with more data in the future.

\begin{turnpage}
\begin{table}[h!]
\begin{center}
\caption{Weighted Mean Parameters of 19-beam Mueller Matrices from 2020 to 2022}
\label{tab:mm}
\begin{tabular}{ccccccccccc}
\hline
\hline
Beam & $\overline{\Delta G}\ (\%)$ & $\sigma_{\overline{\Delta G}}\ (\%)$ & $\overline{\psi}\ (\rm deg)$ & $\sigma_{\overline{\psi}}\ (\rm deg)$ & $\overline{\alpha}\ (\rm deg)$ & $\sigma_{\overline{\alpha}}\ (\rm deg)$ & $\overline{\epsilon}\ (\%)$ & $\sigma_{\overline{\epsilon}}\ (\%)$ & $\overline{\phi}\ (\rm deg)$ & $\sigma_{\overline{\phi}}\ (\rm deg)$\\
\hline
M01 & \phs$0.03 \pm 0.01$ & $0.16 \pm 0.01$ &    $-2.9 \pm 0.1$ & $1.26 \pm 0.05$ &    $-0.03 \pm 0.04$ & $0.32 \pm 0.02$ & $0.141 \pm 0.004$ & $0.028 \pm 0.002$ & $\phn65 \pm \phn2$ &    $18 \pm \phn1$ \\
M02 &    $-3.6 \pm 0.4$ & $0.5 \pm 0.2$     &    $-2 \pm 2$     & $1 \pm 1$     & \phs$1 \pm 1$     & $1.9 \pm 0.5$ & $0.2 \pm 0.1$    & $0.03 \pm 0.06$  & $\phn58 \pm    32$ &    $45 \pm    16$ \\
M03 &    $-0.4 \pm 0.1$ & $1.3 \pm 0.1$     &    $-1 \pm 1$     & $2.0 \pm 0.6$ & \phs$0.2 \pm 0.4$ & $0.4 \pm 0.2$ & $0.17 \pm 0.04$  & $0.06 \pm 0.02$  & $   108 \pm    14$ &    $31 \pm \phn7$ \\
M04 & \phs$1.5 \pm 0.1$ & $1.5 \pm 0.1$     &    $-2 \pm 1$     & $1.3 \pm 0.5$ &    $-0.9 \pm 0.4$ & $2.5 \pm 0.2$ & $0.13 \pm 0.04$  & $0.28 \pm 0.02$  & $   167 \pm \phn9$ &    $50 \pm \phn4$ \\
M05 & \phs$1.0 \pm 0.1$ & $0.85 \pm 0.04$   &    $-1.4 \pm 0.5$ & $0.4 \pm 0.2$ &    $-0.6 \pm 0.2$ & $1.6 \pm 0.1$ & $0.20 \pm 0.02$  & $0.04 \pm 0.01$  & $   109 \pm \phn7$ &    $38 \pm \phn3$ \\
M06 &    $-1.0 \pm 0.2$ & $0.9 \pm 0.1$     & \phs$0 \pm 2$     & $1.1 \pm 0.9$ & \phs$0.4 \pm 0.6$ & $1.2 \pm 0.3$ & $0.09 \pm 0.07$  & $0.09 \pm 0.03$  & $\phn80 \pm    30$ &    $97 \pm    15$ \\
M07 &    $-2.4 \pm 0.3$ & $0.6 \pm 0.1$     & \phs$0 \pm 2$     & $1 \pm 1$     &    $-0.9 \pm 0.7$ & $2.5 \pm 0.4$ & $0.34 \pm 0.09$  & $0.08 \pm 0.05$  & $   109 \pm    13$ &    $29 \pm \phn7$ \\
M08 & \phs$2.7 \pm 0.2$ & $1.0 \pm 0.1$ 	   &    $-1 \pm 1$     & $1.1 \pm 0.5$ & \phs$0.9 \pm 0.4$ & $1.3 \pm 0.2$ & $0.28 \pm 0.05$  & $0.10 \pm 0.02$  & $   136 \pm \phn9$ &    $68 \pm \phn5$ \\
M09 &    $-0.4 \pm 0.1$ & $1.1 \pm 0.1$     &    $-2 \pm 1$     & $1.1 \pm 0.5$ & \phs$0.0 \pm 0.3$ & $1.2 \pm 0.2$ & $0.28 \pm 0.04$  & $0.11 \pm 0.02$  & $   139 \pm \phn8$ &    $14 \pm \phn4$ \\
M10 &    $-0.2 \pm 0.4$ & $2.3 \pm 0.2$     & \phs$0 \pm 4$     & $2 \pm 2$     &    $-0.9 \pm 0.9$ & $1.5 \pm 0.5$ & $0.4 \pm 0.1$    & $0.10 \pm 0.06$  & $   173 \pm    16$ &    $11 \pm \phn8$ \\
M11 &    $-0.6 \pm 0.3$ & $0.2 \pm 0.2$     & \phs$1 \pm 3$     & $2 \pm 2$     & \phs$1 \pm 1$     & $2.6 \pm 0.6$ & $0.34 \pm 0.08$  & $0.01 \pm 0.05$  & $\phn92 \pm    16$ &    $22 \pm \phn9$ \\
M12 &    $-0.9 \pm 0.4$ & $0.4 \pm 0.2$     &    $-1 \pm 3$     & $3 \pm 1$     & \phs$0.5 \pm 0.8$ & $1.5 \pm 0.4$ & $0.3 \pm 0.1$    & $0.30 \pm 0.06$  & $   140 \pm    18$ &    $24 \pm \phn9$ \\
M13 &    $-1.5 \pm 0.2$ & $1.0 \pm 0.1$     &    $-3 \pm 2$     & $2.3 \pm 0.8$ &    $-0.4 \pm 0.5$ & $1.4 \pm 0.3$ & $0.35 \pm 0.06$  & $0.13 \pm 0.03$  & $\phn47 \pm    10$ &    $19 \pm \phn5$ \\
M14 & \phs$1.4 \pm 0.1$ & $0.6 \pm 0.1$     & \phs$0 \pm 1$     & $1.3 \pm 0.5$ & \phs$0.3 \pm 0.4$ & $0.9 \pm 0.2$ & $0.14 \pm 0.04$  & $0.02 \pm 0.02$  & $   136 \pm    16$ &    $31 \pm \phn8$ \\
M15 & \phs$3.0 \pm 0.3$ & $1.2 \pm 0.2$     & \phs$0 \pm 2$     & $1 \pm 1$     & \phs$0.3 \pm 0.7$ & $0.4 \pm 0.4$ & $0.37 \pm 0.07$  & $0.06 \pm 0.04$  & $   166 \pm    14$ & $\phn7 \pm \phn7$ \\
M16 &    $-0.8 \pm 0.2$ & $0.5 \pm 0.1$     &    $-1 \pm 2$     & $1.7 \pm 0.8$ & \phs$0.3 \pm 0.5$ & $1.2 \pm 0.3$ & $0.31 \pm 0.06$  & $0.08 \pm 0.03$  & $   133 \pm    11$ &    $19 \pm \phn5$ \\
M17 &    $-2.3 \pm 0.2$ & $0.8 \pm 0.1$     & \phs$0 \pm 2$     & $1.6 \pm 0.8$ & \phs$0.2 \pm 0.5$ & $1.7 \pm 0.3$ & $0.11 \pm 0.04$  & $0.08 \pm 0.02$  & $\phn85 \pm    24$ &    $44 \pm    12$ \\
M18 & \phs$1.2 \pm 0.3$ & $1.1 \pm 0.2$     &    $-2 \pm 2$     & $2 \pm 1$     &    $-0.4 \pm 0.8$ & $1.8 \pm 0.4$ & $0.2 \pm 0.1$    & $0.06 \pm 0.05$  & $\phn97 \pm    26$ &    $32 \pm    13$ \\
M19 &    $-1.4 \pm 0.2$ & $0.3 \pm 0.1$     &    $-3 \pm 1$     & $1.1 \pm 0.6$ &    $-0.1 \pm 0.4$ & $2.0 \pm 0.2$ & $0.09 \pm 0.05$  & $0.24 \pm 0.02$  & $\phn23 \pm    18$ &    $11 \pm \phn9$ \\
\hline
\end{tabular}
\end{center}
\end{table}
\end{turnpage}

\section{Conclusion} \label{sec:con}
{We performed polarization calibration of the L-band 19-beam receiver using the narrow-band spectral-line backend by observing eight linearly polarized continuum sources within the full illumination pattern of FAST from October 2018 to March 2023, providing a well-sampled data set to study the long-term polarization performance of FAST.} The main results of our polarization calibration are the following: 
\begin{enumerate}
\item Spider observations provide accurate determination of the Mueller matrix parameters of the central beam. After the polarization calibration, the uncertainties in polarization percentage and polarization angle of a source with strong linear polarization emission are between 0.1\% to 0.2\% and about 0.5$^{\circ}$, respectively. The parameters $(\Delta G, \epsilon, \phi)$ show correlations with time in scales from months to years, suggesting relatively frequent polarization calibrations are needed.  
\item We obtain the Mueller matrix parameters of the 18 off-center beams using a combination of OTF observations and spider observations. The consistency of the calibrated fractional Stokes parameters of the 19 beams suggests that the polarization calibration works properly. However, the Mueller matrix parameters of the off-center beams are not as accurate as those of the central beam, making it difficult to study their long-term performance in detail.
\item Although the receiver illuminates different portions of the FAST {reflector surface} toward different sky positions, the Mueller matrix parameters of the central beam at small zenith angles do not show a strong dependence on the dish surface. However, the observations using the eastern side of the {reflector surface} provide parameters $(\alpha, \epsilon, \phi)$ of off-center beams different from those using the western side of the {reflector surface}. The dependence of the off-center Mueller matrix {on the reflector surface} is very likely to be enhanced at large zenith angle which needs to be further measured and analyzed.
\item We provide average parameters of the 19-beam Mueller matrices using the data from 2020 to 2022. The parameters should be applicable to both spectral-line and pulsar observations {at 1420 MHz at small zenith angle with several caveats. After applying those average parameters, on-axis fractional linear polarization measurements $\gtrsim$ 10\% and on-axis fractional circular polarization measurements $\gtrsim$ 1.5\% can be considered high confident detections.
For observations with weak polarization signal from scientific targets, timely polarization calibrations using spider observations, which can provide leakage removal to levels smaller than 0.2\% in linear polarization and 0.1\% in circular polarization, are required.}
\end{enumerate}

\begin{acknowledgments}
\nolinenumbers

We thank JingHai Sun, the PI of FAST project 3053, and Eswaraiah Chakali, the PI of FAST project PT2020\_0078 for providing their data for our work.
The National Radio Astronomy Observatory is a facility of the National Science Foundation operated under cooperative agreement by Associated Universities, Inc.
Lei Qian is supported by National SKA Program of China No.\ 2020SKA0120100, National Nature Science Foundation of China (NSFC) under Grant No.\ 12003047 and 12173053, the Youth Innovation Promotion Association of CAS (id.$\sim$2018075, Y2022027), and the CAS ``Light of West China'' Program.

\end{acknowledgments}

%

\vspace{5mm}
\facilities{FAST:500m}
\software{RHSTK \citep{2022HR}}





\end{document}